\documentclass[twocolumn,british,prl,superscriptaddress,showpacs]{revtex4-1}
\usepackage[letterpaper]{geometry}
\usepackage[dvips]{graphicx}
\usepackage{amsmath}
\usepackage{amssymb}
\usepackage{esint}
\newcommand{\bra}[1]{\langle #1|}
\newcommand{\ket}[1]{|#1\rangle}

\renewcommand{\t}[1]{\textrm{#1}}

\makeatletter
\@ifundefined{textcolor}{}
{%
 \definecolor{BLACK}{gray}{0}
 \definecolor{WHITE}{gray}{1}
 \definecolor{RED}{rgb}{1,0,0}
 \definecolor{GREEN}{rgb}{0,1,0}
 \definecolor{BLUE}{rgb}{0,0,1}
 \definecolor{CYAN}{cmyk}{1,0,0,0}
 \definecolor{MAGENTA}{cmyk}{0,1,0,0}
 \definecolor{YELLOW}{cmyk}{0,0,1,0}
 }

\makeatother

\makeatother

\usepackage{babel}

\begin{document}

\title{Beyond quantum Fisher information: optimal phase estimation with arbitrary a priori knowledge}

\author{Rafa{\l} Demkowicz-Dobrza{\'n}ski}
\affiliation{Faculty of Physics, University of Warsaw, ul. Ho\.{z}a 69, PL-00-681 Warszawa, Poland}

\begin{abstract}
The optimal phase estimation strategy is derived
when partial a priori knowledge on the estimated phase is available. The solution is found with the help
of the most famous result from the entanglement theory: the positive partial transpose criterion.
 The structure of the optimal measurements, estimators and the optimal probe states is analyzed.
The paper provides a unified framework bridging the gap in the literature on the subject which until now dealt almost exclusively with two
extreme cases: almost perfect knowledge (local approach based on Fisher information) and no a priori knowledge
(global approach based on covariant measurements). Special attention is paid to a
natural a priori probability distribution arising from a diffusion process.
\end{abstract}


\maketitle


Quantum states may be employed
to improve the precision of measurements without committing additional resources such as energy or time
 \cite{Bollinger1996}. 
 A paradigmatic model is the estimation of relative phase delay between two arms of a Mach-Zehnder interferometer
 \cite{Caves1981}. 
When the interferometer is fed with a light pulse prepared in a coherent state, the precision of phase estimation
scales as $\delta \varphi \propto 1/\sqrt{N}$, where $N$ is the mean number of photons in a pulse.
On the other hand, by preparation of $N$ photons, entering the
two input ports of the interferometer, in a appropriate quantum
state, a quadratic precision enhancement may be achieved  leading to
$\delta \varphi \propto 1/N$ referred to as the Heisenberg limit \cite{Giovannetti2004}.

These idealized results need to be contrasted with a
more realistic ones when environmental noise and experimental imperfections are taken into account
\cite{Huelga1997,Dorner2008}. 
For optical implementations the most relevant disruptive factor is the photon loss.
For large $N$, the precision of optimal phase estimation approximates to \cite{Knysh2010, Kolodynski2010}
$\delta \varphi \approx \sqrt{1-\eta}/\sqrt{\eta N}$, where
$\eta$ is the overall power transmission of an interferometer
(including scattering, reflections, detector efficiencies etc.).
The Heisenberg scaling is asymptotically lost and the quantum precision enhancement amounts to the $\sqrt{1 -\eta}$ factor.
Although asymptotically the prospect of quantum enhanced metrology may look bleak,
still for moderate $N$ 
the advantage of using entangled states may be very significant \cite{Dorner2008}.

Interestingly, the above asymptotic formula was obtained independently using two different approaches.
The first, \emph{local approach}, aims at maximizing the estimation sensitivity
to small phase variations around an a priori known value $\varphi=\varphi_0$. The main theoretical tool is
the \emph{quantum Fisher information} (QFI), $F_Q$, which defines an asymptotically achievable quantum Cram{\'e}r-Rao (CR)
bound \cite{Braunstein1994}
on estimation precision $\delta \varphi \geq 1/\sqrt{F_Q}$. The second, \emph{global approach}, assumes
no a priori knowledge
on the phase --- a priori probability distribution is uniform over the $[-\pi,\pi)$ region --- and makes use of the phase shift symmetry
to restrict the class of measurements to the covariant ones \cite{Holevo1982}. 
It is quite intuitive that in asymptotic regime the role of a priori knowledge should be negligible,
since in principle a minute fraction of resources might be committed to preestimate the phase and compensate for the lack
of a priori knowledge.

For finite $N$, however, a priori knowledge may strongly influence both the optimal precision and
the estimation strategy itself. Since the main prospects
for applications of quantum enhanced metrology lay in the regime of moderate values of $N$,
the amount of a priori knowledge may play a key role in designing the optimal estimation
schemes. Moreover, the problem is highly relevant for feedback
estimation schemes, where in each iteration one should make the
optimal use of the information gained in the previous steps. This paper provides a direct way to find the optimal estimation schemes for an arbitrary form of
a priori knowledge.
As a result, one can also study the transition from
 \emph{global} to  \emph{local} approaches by tuning the a priori probability distribution
 from uniform, $p(\varphi)=1/2\pi$, to peaked $p(\varphi)\approx \delta(\varphi -\varphi_0)$
  (see \cite{Durkin2007} which is one of very few papers approaching the problem of phase estimation in the intermediate regime).

Let $\mathcal{H}$ be an $N+1$ dimensional Hilbert space, and let $\ket{n} \in \mathcal{H}$, $n=0\dots N$,
denote an eigenbasis of the phase shift operator $U_\varphi$, 
$U_\varphi \ket{n} = \exp(-i n \varphi) \ket{n}$. Thinking in terms of an $N$ photon state inside a Mach-Zehnder interferometer,
  $\ket{n}$ denotes a state in which $n$ photons travel through the upper and $N-n$ photons through the lower arm,
  while the upper arm is delayed by $\varphi$ with respect to the lower one. Let $\rho=\sum_{nm} \rho^n_m \ket{n}\bra{m}$
  be the probe state, which under the phase shift becomes $\rho_\varphi = U_\varphi \rho U^{\dagger}_\varphi$.
  Note, that $\rho$ may represent a decohered pure probe state, provided the decoherence process commutes with the phase shift operation, which
  is the case for the most relevant decoherence mechanisms such as photon loss in optical and
  dephasing or damping in atomic systems \cite{Huelga1997, Dorner2008}.

 The estimation strategy is defined by a POVM measurement
 $\Pi_{\tilde{\varphi}}$, $\int_{-\pi}^{\pi} \t{d} \tilde{\varphi}\  \Pi_{\tilde{\varphi}} = \openone$, $\Pi_{\tilde{\varphi}} \geq 0$,
 where the measurement outcome $\tilde{\varphi}$ is at the same time the estimator of the phase. Hence,
$\t{Tr}(\rho_\varphi \Pi_{\tilde{\varphi}})$ is the conditional probability of estimating $\tilde{\varphi}$ provided the true phase is $\varphi$.
 Taking into account the a priori probability distribution $p(\varphi)$, the average cost of the estimation strategy reads:
 \begin{equation}
 \label{eq:cost}
 \overline{C} = \int_{-\pi}^{\pi} \t{d}\varphi \t{d}\tilde{\varphi} \, p(\varphi) \t{Tr}(\rho_\varphi \Pi_{\tilde{\varphi}}) C_{\varphi,\tilde{\varphi}},
 \end{equation}
where $C_{\varphi,\tilde{\varphi}}$ is the cost for guessing $\tilde{\varphi}$, while the true value was $\varphi$.
The $C_{\varphi,\tilde{\varphi}}=(\varphi-\tilde{\varphi})^2$
cost function is only appropriate for narrow distribution around $\varphi=0$ since it does not respect
periodic conditions $C_{\varphi+2\pi,\tilde{\varphi}}$. We choose 
 $C_{\varphi,\tilde{\varphi}} = 4 \sin^2\frac{{\varphi-\tilde{\varphi}}}{2}$,
as it is the simplest function (with the least number of Fourier components)
that approximates the variance for narrow distributions \cite{Berry2000}, and denote its average by $\widetilde{\delta^2 \varphi}$.

Consider an auxiliary two dimensional Hilbert space $\mathcal{H}_A$,
where we 
define $\ket{\varphi}=(\ket{0}+\exp(-i \varphi)\ket{1})/\sqrt{2}$.
Notice that 
\begin{equation}
C_{\varphi, \tilde{\varphi}}= 
 4\left[1- \t{Tr}\left(\ket{\varphi}\bra{\varphi}  \ket{\tilde{\varphi}}\bra{\tilde{\varphi}}\right)\right].
 \end{equation}
Substituting the above formula to Eq.~(\ref{eq:cost}), making use of the fact that $\t{Tr}(AB)\t{Tr}(CD)=\t{Tr}[(A\otimes C) (B\otimes D)]$ and
inserting both integrals under the trace we finally arrive at:
\begin{equation}
\label{eq:costvar}
\widetilde{\delta^2 \varphi}=\overline{C}=4\left(1-F\right), \quad F=\t{Tr}\left(R M\right)
\end{equation}
where
\begin{eqnarray}
\label{eq:rhoform}
R&=& \int_{-\pi}^\pi \t{d}\varphi \, p(\varphi) \ket{\varphi}\bra{\varphi} \otimes \rho_\varphi,\\
M&=& \int_{-\pi}^\pi \t{d}\tilde{\varphi} \,\ket{\tilde{\varphi}}\bra{\tilde{\varphi}}\otimes \Pi_{\tilde{\varphi}}.
\label{eq:piform}
\end{eqnarray}

Assuming the probe state $\rho$ is given, the problem of finding the optimal estimation strategy amounts to
maximizing the fidelity $F$, over operators $M$ which are of the form (\ref{eq:piform}).
The structure of $M$ is analogous to the structure of a separable state, with an additional constraint resulting from
the POVM completeness condition: $\t{Tr}_A M = \int_{-\pi}^\pi \t{d} \varphi \Pi_{\varphi} = \openone$.
This observation has been employed in \cite{Navascues2008} to cast
the problem of optimal pure state reconstruction into
maximization of linear functional over separable states, which in principle may be
solved numerically using semi-definite programming.
Surprisingly,  as demonstrated below, the solution to the problem considered in this paper is found analytically,
without resorting to semi-definite programming, and just with the help of simple positive partial transpose (PPT)
necessary criterion for separability \cite{Peres1996}.



Let us write the a priori distribution as $p(\varphi)=\frac{1}{2\pi}\sum_{k=-\infty}^\infty p_k \exp(i k \varphi)$ where
$p_k^* = p_{-k}$ guarantees $p(\varphi) \in \mathbb{R}$ and $p_0=1$ assures normalization.
Both $R$ and $M$ may be written using a block form respecting
the structure of the tensor product $\mathcal{H}_A \otimes \mathcal{H}$:
\begin{equation}
    R=
    \left(\begin{array}{cc}
           R_0^0 & R^0_1 \\
           R^1_0 & R^1_1
    \end{array}
    \right),
    \
    M=
    \left(\begin{array}{cc}
           M^0_0 & M^0_1 \\
           M^1_0 & M^1_1
    \end{array}
    \right)
\end{equation}
where $R^i_j = {_A\bra{i}} R \ket{j}_A$, $M^i_j = {_A\bra{i}} M \ket{j}_A$.
Straightforward calculation of $R^i_j$ using Eq.~(\ref{eq:rhoform}) yields
\begin{equation}
(R^i_i)^n_m=\tfrac{1}{2}\rho^n_m p_{n-m}, \ (R^0_1)^n_m= (R^1_0)^{m *}_n = \tfrac{1}{2}  \rho^n_m p_{n-m+1}.
\end{equation}
The completeness constraint $\t{Tr}_A M=\openone$, implies $M^0_0+M^1_1=\openone$.
Using the fact that $R^0_0=R^1_1$ and
$\t{Tr} R^i_i =1/2$, we get $F=\frac{1}{2} + 2 \Re\left[\t{Tr}(R^{1}_0 M^0_1)\right]$.

\emph{Main result.} The cost of the optimal phase estimation strategy reads
\begin{equation}
\label{eq:optimalcost}
\widetilde{\delta^2 \varphi} = 4\left(\tfrac{1}{2} - \|R^1_0\|_1\right),
\end{equation}
where $\|A\|_1$ denotes the trace norm of a matrix. The optimal estimation strategy
itself is given by
\begin{equation}
\label{eq:optimalpi}
M = \sum_{k=0}^{N} \ket{\varphi_k}\bra{\varphi_k} \otimes \ket{\psi_k} \bra{\psi_k}
\end{equation}
where $\exp(-i \varphi_k)$, $\ket{\psi_k}$ are eigenvalues and eigenvectors of
$U=V_R U_R^\dagger$, while  $V_R$, $U_R$ are unitaries appearing in the singular value decomposition (SVD)
$R^1_0=U_R \Lambda_R V_R^\dagger$.
Notice that Eq.~(\ref{eq:optimalpi}) provides an explicit recipe for implementation
of the optimal estimation strategy. The optimal measurement is a projective measurement on
basis $\ket{\psi_k}$, with  $\varphi_k$ being the estimated phase given a measurement result $k$.

\begin{figure*}[t]
\includegraphics[width=1\textwidth]{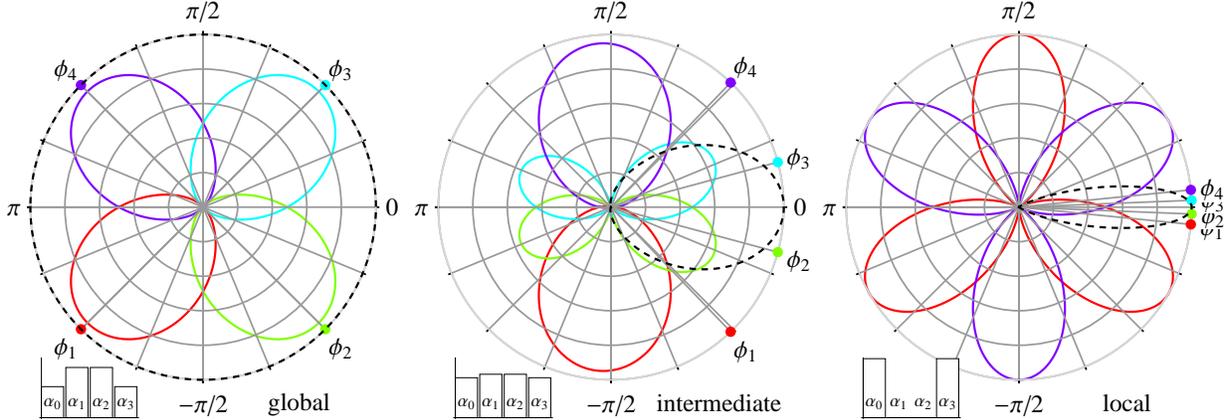}
\caption{(color online) Optimal phase estimation strategies for $N=3$ photon probe states $\ket{\psi}$
and different degrees of a priori knowledge $p_t(\varphi)$ (black, dashed --- rescaled so that $p_t(0)=1$) [see Eq.~(\ref{eq:diff})]
 for $t=20$ (global regime), $t=0.2$ (intermediate regime), $t=0.02$ (local regime) respectively.
 $\varphi_k$ is a phase that is estimated once a measurement result $k$ is obtained, while the corresponding curve
 depicts $p(k|\varphi)=|\bra{\psi_k}U_\varphi \ket{\psi}|^2$ --- conditional probability that measurement outcome $k$
 is obtained if the true phase is $\varphi$. Insets in the bottom left corners illustrate parameters $\alpha_k$ of the optimal
 probe state. In the local regime only two
 measurement outcomes are relevant and the optimal state is the N00N state --- a result
 known from the quantum Fisher information approach.
 }
 \label{fig:structure}
\end{figure*}
\emph{Proof.} Let $M^\prime$ represent the optimal measurement strategy,
and define $M^{\prime \prime}$ such that off-diagonal blocks remain the same while the diagonal
blocks are interchanged: $M^{\prime \prime 0}_0 = M^{\prime 1}_1$,  $M^{\prime \prime 1}_1 = M^{\prime 0}_0$.
$M^\prime$ is of course positive semi-definite, but this is not guaranteed for $M^{\prime \prime}$.
Notice, however, that $M^{\prime \prime} = W M^{\prime T_A} W^\dagger$, where
$W=\left(\begin{array}{cc} 0 & \openone \\ \openone  & 0 \end{array} \right)$ is a unitary operation
 and $T_A$ is the partial transposition with respect to the $A$ subsystem.
Since $M^{\prime}$ is separable, then by the PPT criterion  $M^{\prime T_A} \geq 0$, and hence also
$M^{\prime \prime} \geq 0$.

$F$ depends only on the off-diagonal blocks via $\Re\left[\t{Tr}(R^{1}_0 M^0_1)\right]$.
Therefore, $M^{\prime \prime}$ provides the same fidelity as $M^{\prime}$.
Keeping fidelity the same we construct
\begin{equation}
M =(M^\prime+M^{\prime \prime})/2 = \left(\begin{array}{cc}
  \openone/2 & M^{\prime 0}_1 \\ M^{\prime 0 \dagger}_1 &\openone/2  \end{array} \right).
\end{equation}
Positive semi-definiteness of $M$ implies that $M^{\prime 0 \dagger}_1 M ^{\prime 0}_1 \leq \openone/4$,
hence all singular values of $M^{\prime 0}_1$ are no greater than $1/2$.

Let $R^1_0 = U_R \Lambda_R V^\dagger_R$, $M^0_1 = U_M \Lambda_M V^\dagger_M$ be SVDs.
Thanks to $\Lambda_M \leq \openone/2$, and the fact that
 $|\t{Tr} \left( U \Lambda \right)| \leq \t{Tr}\Lambda$
  the following chain of inequalities holds:
\begin{multline}
\Re\left[\t{Tr}\left(R^{1}_0 M^0_1\right)\right] \leq 
 \left|\t{Tr}\left( U_R \Lambda_R V^\dagger_R U_M \Lambda_M V^\dagger_M\right)\right| \leq \\
\leq \frac{1}{2} \t{Tr}\left(\Lambda_R\right) =
 \frac{1}{2} \|R^1_0\|_1.
\end{multline}
All inequalities are saturated for $M^0_1 = U/2$, where $U=V_R U^\dagger_R$.
Using the eigen decomposition
$U=\sum_{k=0}^N \exp(- i \varphi_k) \ket{\psi_k}\bra{\psi_k}$
we arrive at Eq.~(\ref{eq:optimalpi}). $\blacksquare$

In the following examples, we choose a prior $p(\varphi)$ that is a solution of the
diffusive evolution on the $[-\pi,\pi)$ interval with periodic
boundary conditions and initial $\delta(\varphi)$ distribution:
\begin{equation}
\label{eq:diff}
p_t(\varphi)=\frac{1}{2\pi}\left(1 + 2 \sum_{n=1}^\infty \cos(n \varphi) e^{- n^2 t}\right).
\end{equation}
This naturally approximates the gaussian distribution on the interval, with $t=0$ corresponding to $\delta(\varphi)$
 and $t=\infty$ to uniform distribution:

For the sake of clarity, in the following examples we consider only the decoherence-free phase estimation, where
the probe state is pure $\rho=\ket{\psi}\bra{\psi}$, $\ket{\psi}=\sum_{n=0}^N \alpha_n \ket{n}$ (as stressed before
the method can be applied to more general scenarios).
In this case, $(R^1_0)^n_m = \tfrac{1}{2}\alpha_n \alpha_m^* \exp[-(m-n+1)^2 t]$, and the optimal
probe state corresponds to the choice of $\alpha_n$ such that the trace norm of the above matrix
is maximized. For $t=\infty$ (global approach),
$ (R^1_0)^n_m = \frac{1}{2}\alpha_n \alpha_{n-1} \delta_{n-1,m}$,
$\|R^1_0 \|_1=\frac{1}{2}\sum_{n=0}^{N-1} |\alpha_n| |\alpha_{n+1}|$ which is maximized
 by the Berry-Wiseman (BW) states \cite{Berry2000}.
For general $t$ there is no analytical formula for the trace norm as a function of $\alpha_n$ and
one needs to look for the optimal probe state numerically.

To facilitate the numerical search, 
one may implement an efficient iterative procedure. In the $i$-th step starting with a pure probe state
$\rho^{(i)}=\ket{\psi^{(i)}}\bra{\psi^{(i)}}$, one calculates the optimal
estimation strategy $M^{(i)}$ --- as shown above this requires only a single run of SVD algorithm.
Having found $M^{(i)}$, one uses Eq.~(\ref{eq:costvar}) and looks for the $\ket{\psi^{i+1}}$ maximizing
$F=\t{Tr}(R^{(i+1)} M^{(i)})= \bra{\psi^{(i+1)}} \mathfrak{M}^{(i)} \ket{\psi^{(i+1)}}$ where
\begin{equation}
\mathfrak{M}^{(i)} =\int \,\t{d} \varphi p(\varphi) \bra{\varphi} \openone \otimes U_\varphi^\dagger \ M^{(i)}\  \openone
\otimes U_{\varphi}   \ket{\varphi}.
\end{equation}
Optimal $\ket{\psi^{i+1}}$ is the eigenvector corresponding to the maximum eigenvalue of the matrix $\mathfrak{M}^{(i)}$.

The structure of the optimal estimation strategy and the optimal probe state is depicted in Fig.~\ref{fig:structure}
for $N=3$ and three different a priori distributions. Notice, how, with the increasing a priori knowledge, the optimal probe state
evolves to the N00N state \cite{Bollinger1996}, which is the optimal solution of the QFI approach.
The periodic ($2\pi/N$) structure of the conditional probabilities $p(k|\varphi)$, visible in the local regime, clearly
reminds of the fact that this estimation strategy is useless unless the prior is highly peaked since otherwise
there is strong ambiguity in using the measurement result to estimate the phase.

It is worth mentioning, that although both the local approach discussed in this paper (corresponding to the limit $t \rightarrow 0$),
and the QFI approach yield the same optimal probe states and the same
optimal measurements, they in general yield different estimation precisions.
While $F_Q$ is an extremely useful tool, it just provides a CR bound
on the achievable estimation precision
$\delta \varphi \geq 1/\sqrt{F_Q}$. This bound, in general, cannot be saturated by an estimator
based on the results of a single measurement. Only in the asymptotic limit of infinitely many repetitions of the experiment
one may construct a max-likelihood estimator which saturates the bound.
The approach presented in this paper, on the other hand, gives an operationally meaningful answer for single  shot
estimation procedure. Moreover, when $t \rightarrow 0$, then $\delta \varphi \rightarrow 0$ by the obvious
fact that that the phase is known perfectly. Hence, for $t$ small enough the CR bound is violated
(this is no contradiction since our estimator is not locally unbiased \cite{Helstrom1976}).

\begin{figure}[t]
\includegraphics[width=1\columnwidth]{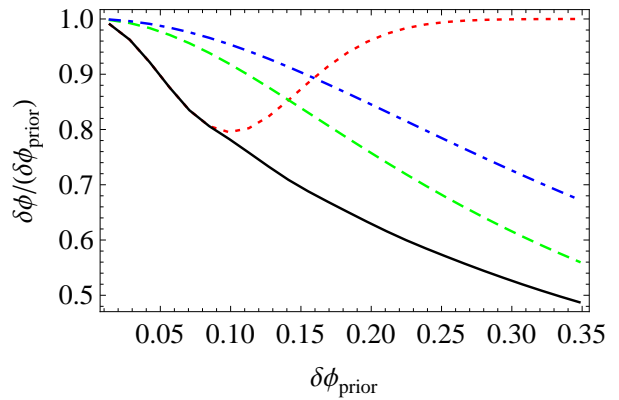}
\caption{(color online) Relative reduction of the uncertainty $\delta \varphi$ with respect
to $\delta \varphi_{\t{prior}}$ after the optimal estimation procedure is applied
to the $N=10$ optimal probe state (black, solid); N00N state (optimal in the local regime)
 (red, dotted);
BW state (optimal in the global regime)
(green, dashed); ,,classical'' state with $\alpha_k^2=\binom{N}{k}$ (blue, dash-dotted).
} \label{fig:gain}
\end{figure}
Performance of the optimal estimation strategy for $N=10$ and different probe states is depicted in Fig.~\ref{fig:gain}
as a function of a priori uncertainty $\delta \varphi_{\t{prior}}=\sqrt{\int \t{d}\varphi \, 4 \sin^2(\varphi/2) p_t(\varphi)}$.
N00N states are optimal up to a threshold (which scales as $1/N$) above which
they become useless due to $2\pi/N$ phase estimation ambiguity. The optimal states clearly demonstrate their superiority over
the BW states for moderate $\delta \varphi_{\t{prior}}$ and lose
their advantage with the increasing prior ignorance on the phase value.

In summary, the problem of optimal phase estimation with arbitrary a priori knowledge has been solved analytically,
allowing to investigate the regime of estimation not accessible via neither the QFI approach nor
covariant measurements. Optimal measurements and estimators have been explicitly constructed so that
it is immediate to apply the results to any practical phase estimation problem.

I would like to thank Konrad Banaszek for constant support and many valuable suggestions.
This research was supported by the European Commission under
the Integrating Project Q-ESSENCE and the Foundation for Polish Science under the TEAM programme.

%

\end{document}